\documentclass[a4paper, 10pt, conference]{ieeeconf}      % Use this line for a4 paper

\IEEEoverridecommandlockouts                              % This command is only needed if 
\usepackage[noadjust, nobreak]{cite}

\usepackage{graphicx}
\graphicspath{{./graphics/}}
\usepackage[utf8]{inputenc}
\usepackage{multirow}

\usepackage{todonotes}
\usepackage{xcolor}

\usepackage{nicefrac}

\usepackage{amsmath}
\usepackage[noadjust]{cite}

\usepackage{gensymb}

\title{\LARGE \bf Driver Assistance for Safe and Comfortable On-Ramp Merging Using Environment Models Extended through V2X Communication and Role-Based Behavior Predictions}

\author{Lucas~Eiermann\textsuperscript{\includegraphics[scale=0.4]{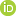}}, Florian~Wirthmüller\textsuperscript{\includegraphics[scale=0.4]{orcidlogo.png}}, Kay~Massow\textsuperscript{\includegraphics[scale=0.4]{orcidlogo.png}}, Gabi Breuel and Ilja Radusch
\thanks{L. Eiermann, F. Wirthmüller and G.Breuel are with Mercedes-Benz AG, Böblingen, Germany, E-Mail: \{first\_name.last\_name\}@daimler.com}
\thanks{F. Wirthmüller is with the Institute of Databases and Information Systems (DBIS), Ulm University, Ulm, Germany}
\thanks{K. Massow and I. Radusch are with Daimler Center for Automotive Information Technology Innovations (DCAITI) at TU Berlin, Berlin, E-Mail: \{first\_name.last\_name\}@tu-berlin.de}
\thanks{I. Radusch is with Fraunhofer FOKUS, Berlin}
\thanks{ORCID: \href{https://orcid.org/0000-0001-9538-5991}{https://orcid.org/0000-0002-9732-2561};\newline
\href{https://orcid.org/0000-0002-3760-5762}{https://orcid.org/0000-0002-9732-2561};\newline 
\href{https://orcid.org/0000-0002-3760-5762}{https://orcid.org/0000-0002-3760-5762}}%\thanks{\copyright~2020 IEEE. Personal use of this material is permitted. Permission from IEEE must be obtained for all other uses, in any current or future media, including reprinting/republishing this material for advertising or promotional purposes, creating new collective works, for resale or redistribution to servers or lists, or reuse of any copyrighted component of this work in other works.}
}

\usepackage{url}
\usepackage{hyperref}

\IEEEoverridecommandlockouts
\pubid{978-1-7281-9080-8/20/\$31.00  \copyright~2020 IEEE}

\hypersetup{bookmarks={false}}

\begin{document}

\setlength{\abovedisplayskip}{2pt}
\setlength{\belowdisplayskip}{3pt}

\maketitle
\pagestyle{empty}

%%%%%%%%%%%%%%%%%%%%%%%%%%%%%%%%%%%%%%%%%%%%%%%%%%%%%%%%%%%%%%%%%%%%%%%%%%%%%%%%

\begin{abstract}
Modern driver assistance systems as well as autonomous vehicles take their decisions based on local maps of the environment. These maps include, for example, surrounding moving objects perceived by sensors as well as routes and navigation information. Current research in the field of environment mapping is concerned with two major challenges. The first one is the integration of information from different sources e.\,g. on-board sensors like radar, camera, ultrasound and lidar, offline map data or backend information. The second challenge comprises in finding an abstract representation of this aggregated information with suitable interfaces for different driving functions and traffic situations. 
To overcome these challenges, an extended environment model is a reasonable choice.

In this paper, we show that role-based motion predictions in combination with v2x-extended environment models are able to contribute to increased traffic safety and driving comfort. Thus, we combine the mentioned research areas and show possible improvements, using the example of a threading process at a motorway access road. Furthermore, it is shown that already an average v2x equipment penetration of 80\,\% can lead to a significant improvement of 0.33\,{$\nicefrac{m}{s^2}$} of the total acceleration and 12\,m more safety distance compared to non v2x-equipped vehicles during the threading process.
\end{abstract}

%of all road users equipped with a state-of-the-art sensor setup

%\todo{abstract: das mit den maps ist verwirrend - einmal bedeutet es abbildung und einmal karte --> versuch das sprachlich etwas zu trennen}

%%%%%%%%%%%%%%%%%%%%%%%%%%%%%%%%%%%%%%%%%%%%%%%%%%%%%%%%%%%%%%%%%%%%%%%%%%%%%%%%
\section{INTRODUCTION}
In recent years, great progress in the field of autonomous driving has been achieved. Latest technologies such as computer vision, AI technologies and scalable high-performance hardware systems have contributed to accomplish that. Nevertheless, it still remains challenging to generate a robust perception of safety-relevant information. As current research not only focuses on solving the actual driving task but also on increasing comfort and safety in particularly demanding situations this challenge is even aggravated. Further important goals to be achieved with the help of automated vehicles are increasing traffic efficiency and reducing traffic jams. To achieve these ambitious goals, a combination of complementary sensor systems is required. This allows an optimal ego-perspective of the existing traffic situation to be depicted. In daily traffic, however, situations arise in which automated systems cannot react optimally due to limitations of their on-board sensor equipment. Especially, in these situations, automated systems are highly dependent on cooperating and communicating with others to obtain an accurate impression of their surrounding.

\begin{figure}[t!]
\centering\includegraphics[width=0.28\textwidth]{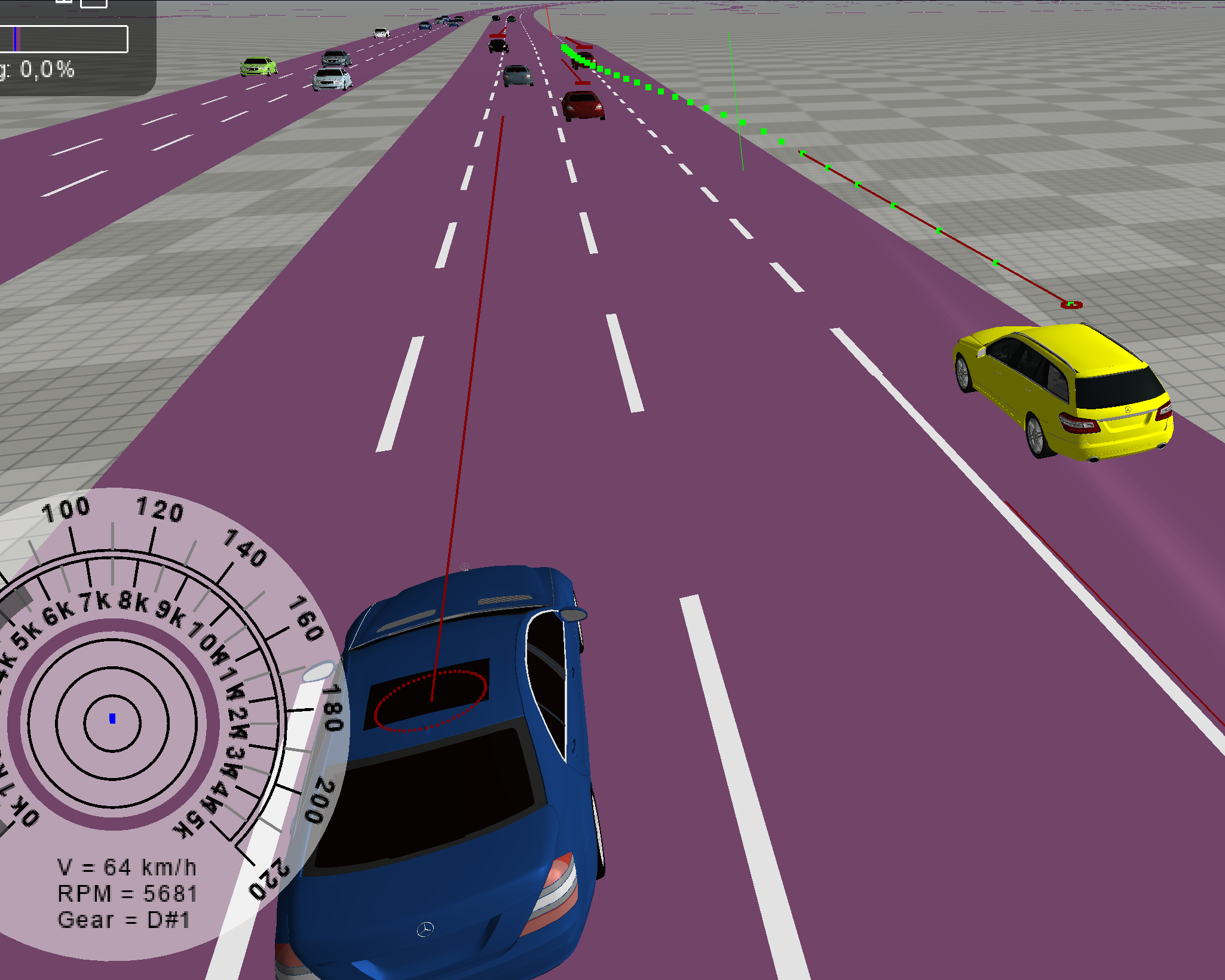}\caption{PHABMACS 3D simulation with v2x-extended envirionment (red lines and circles) and planned ego-trajectory (green points).}\label{fig:phab_es_visu}\vspace{-0.5cm}
\end{figure} 

\pubidadjcol

Threading operations are one of the most difficult maneuvres in road traffic. In contrast to lane changes, which can be carried out locally in a largely flexible manner, they inevitably occur in places where the number of lanes is reduced. Applications can be construction-related lane subtractions, or the more widespread example of accessing a motorway. While doing so, drivers look for gaps in the flowing traffic on the main lane to merge in. The driver has to solve different tasks at the same time: 
\begin{itemize}
\item Adapting his speed to the flowing traffic 
\item Considering surrounding vehicles
\item Respecting the end of the track
\item Finding suitable gaps in terms of size and accessibility and making a selection for one of them
\item Predicting the dynamic development of the gap (speeds of the limiting vehicles, often also more or less clearly recognizable intentions of the respective drivers with regard to cooperation) 
\end{itemize}

A cooperative assistance function can provide advantageous support in solving this demanding driving task. Through the aspect of v2x shared objects, gap sizes can be precisely determined. Besides, reliable forecasts of their development can be made by evaluating the driving data of the vehicles involved. Furthermore, the risk of vehicles being "overlooked" is minimized by incorporating on-board sensors. \\

In this article, we show the significant comfort and safety gain being achieved through using v2x techniques, especially when merging at motorway entrances. The term v2x describes a communication in one way or another between a vehicle and another vehicle (v2v) or an infrastructure element (v2i) such as a traffic light or stationary sensor. The presented approach is based on an extended environment model, which was implemented with the two message types Cooperative Awareness Message (CAM) and Collective Peception Message (CPM)\footnote{introduced in detail in \autoref{sec:rel_work}}. The fusion of the messages takes place decentrally in the vehicle computer. Based on the fused information, an extended environment model is created, which determines the positions and variances of the surrounding vehicles by means of a Kalman filter. These information are complemented by the output of a role-based neural network position prediction module, which estimates the behavior of surrounding traffic participants and the developement of traffic gaps. Thus, significant improvements regarding comfort and safety are achieved. \autoref{fig:phab_es_visu} shows the extended environment model and the planned trajectory within the used 3D simulation framework PHABMACS (PHysics Aware Behavior Modeling Advanced Car Simulator). The v2v communication facility and the object sensors are modelled as motivated in \cite{Massow.2018 }.

The remainder of this work is structured as follows: \autoref{sec:rel_work} presents the required message types as well as relevant publications in the area of v2x communication and behavior prediction. \autoref{sec:sys_overview} describes the system's structure and the integration of different modules . \autoref{sec:imp_and_sil_testing} describes the implementation of the functions. In \autoref{sec:eval_and_results} the approach is evaluated and discussed. Finally, \autoref{sec:conclusion} draws a conclusion and gives a short outlook on future work.

\section{STATE OF THE ART}\label{sec:rel_work}
Cooperative Awareness (CA) is a mandatory and security relevant basic application for v2x communication and is already completely standardised in EN 302-637-2 \cite{EuropeanTelecommunicationsStandardsInstitute.2013} by ETSI for the European area. Each station in an ITS-G5 network periodically distributes this message to its immediate environment. The CA service contains information from the station about its position, speed and direction of travel. By receiving this information, each station increases its own perception of its immediate surroundings.
In this paper we used the Cooperative Awareness Message (CAM) to fill the extended environment model and thus provide merged object lists to the v2x-enabled vehicles. The CA messages can be used to transmit information like positions, orientation angles and speeds to the environment. In addition, sending the message can also transmit information about the vehicle ID, dimensions, vehicle type and weight to the environment. 

\begin{figure}[t!]
\centering\includegraphics[height=4.7 cm]{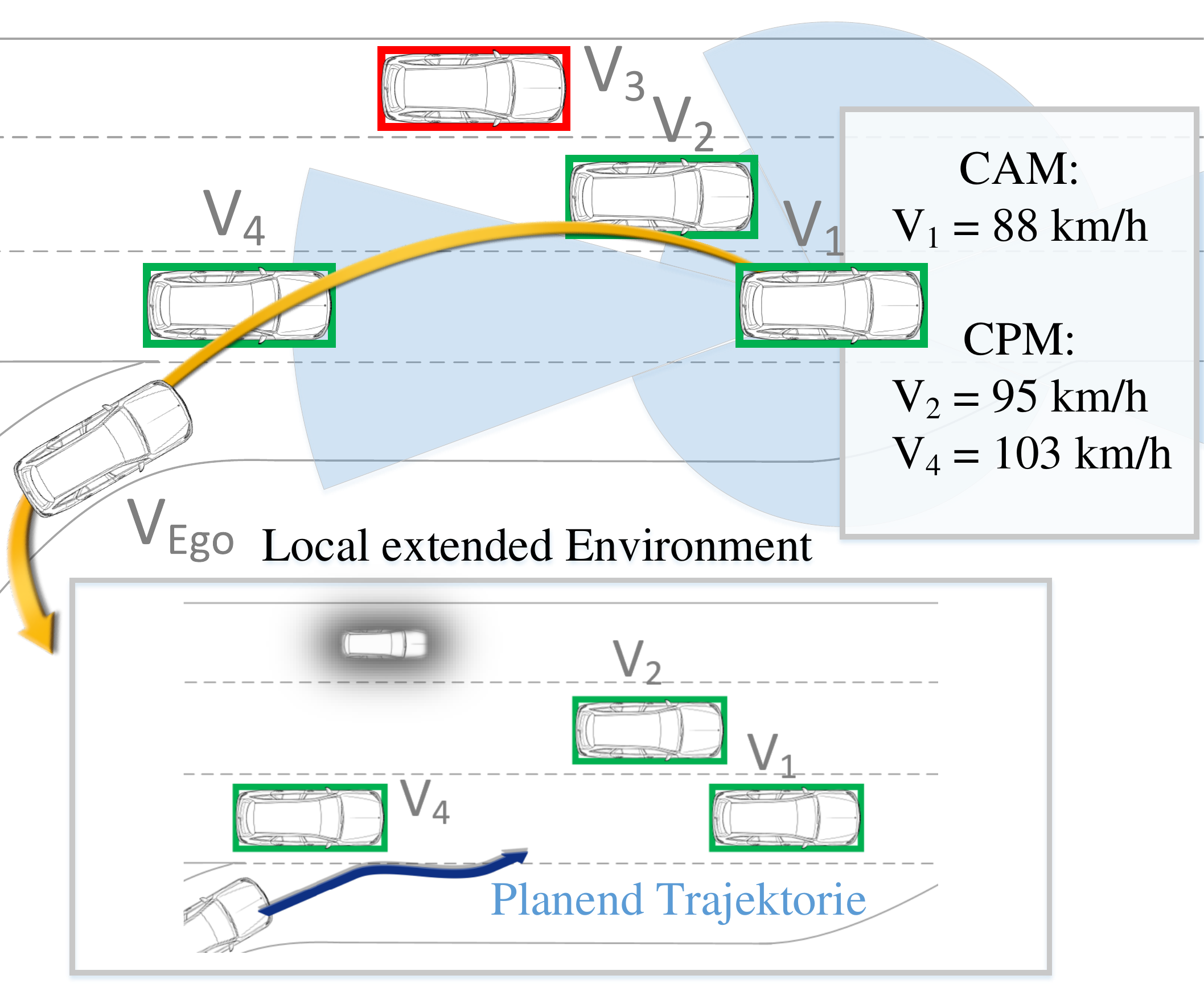}\caption{Overview of the environment model extended with CAM and CPM. Vehicle $V_2$ sends the detected vehicle ($V_2$, $V_4$) positions to the ego vehicle $V_{Ego}$.}\label{fig:ext_env_overview}\vspace{-0.5cm}
\end{figure}

As a supplement to the CAM, the system was equipped with collective perception (CP) in our investigations. CP describes the information gain by communicating the sensory perceived objects of the transmitter to its surrounding vehicles. For this purpose, the standardization committee defined a new messag called Collective Peception Message (CPM). Thus, perception messages can be used to fill the extended environment model with additional environment objects that may not have v2x functionality. This is an enormous gain, as it ensures large coverage areas. \autoref{fig:ext_env_overview} illustrates the extension of the environment model through exchanging the described messages for a demonstartive example. 

With the above-mentioned messages, v2x communication offers the opportunity of informing vehicles in demanding or dangerous situations about the environment. Therefore, the ego-vehicle can adapt its planned behavior in order to increase safety, comfort and traffic flow efficiency. Another major advantage is that the approach can be applied and scaled to future mixed traffic with vehicles equipped with different assistance and automation systems.

Existing communication protocols such as ETSI ITS-G5 \cite{EuropeanTelecommunicationsStandardsInstitute.2013} and SAE J2735 \cite{SAEInternational.2016} focus on human-centric vehicle operation. The next generation of this communication should focus on a cooperative and automated driver function to enable autonomous driving under difficult conditions. 
In addition, automated vehicles can share trajectories and other information about their vehicle conditions. This information increases the accuracy of the prediction interval. For short term maneuvers it is possible to exchange planned trajectories for maneuver coordination \cite{Dixit.2018}. 

Sawade \cite{Sawade.2018} proposes his approach to negotiate driving maneuvers with role-based distributed state machines. The article shows that role-based communication cooperation is very well suited for highly and fully automated vehicles. Even under adverse conditions, such as those encountered in wireless ad hoc networks, good results are shown in maneuver management. In this work we transfer Sawades role-based communication setup to a role-based prediction. This allows mixed traffic with automated and non-automated vehicles. Vehicles that are v2x-enabled communicate their role over the network. The roles of other vehicles, which are not equipped with communication technology, are identified via a neural network. This improves the prediction of road users and enables interaction with other road users. This is a very important functionality especially for on-ramps.

\cite{Eiermann.2020} shows that role-based negotiation is also suitable for complex on-ramp merges. Furthermore, a system overview of an existing maneuver planner as used in this paper is presented. In addition, this paper focuses on the environment model as well as the role recognition of non-v2x capable vehicles. This is to meet the challenge of dealing with different degrees of automation, i.\,e. mixed traffic. For this purpose an extended environment model is used to improve the role recognition with a neural network.

The extended environment model is realized by using an extended Kalman filter, as shown in \cite{Becker.2000, Becker.58Oct.1999}. These approaches aim at optimizing sensor fusion for multiple sensor arrays in autonomous vehicles. This is done with fusion algorithms that associate different object features. Using a Kalman filter with various measurements inputs, a position is determined. The computing time for this fusion algorithm is twice as long as that of the Kalman filter.

In \cite{Dickmann.2015} the concept of collective perception was investigated in connection with v2x applications. The focus was put on the message structure and the resulting network load. The sensor information of a traffic participant is sent to a second vehicle in a presented protocol procedure, taking into account local sensors and v2x information. This enables the vehicles to react to critical traffic situations in time. In this thesis the framework as well as the message was realized on two automated vehicles. The benefit was analyzed in a scenario for avoiding obstacles. It was shown that the resulting available reaction time to bypass the obstacle on the route increases significantly. In the presented scenario an average reaction time for vehicles without v2x connectivity (i.\,e. with local sensors only) of 3.1\,s on average was measured to avoid collisions.

\begin{figure}[t!]
\centering\includegraphics[width=0.43\textwidth]{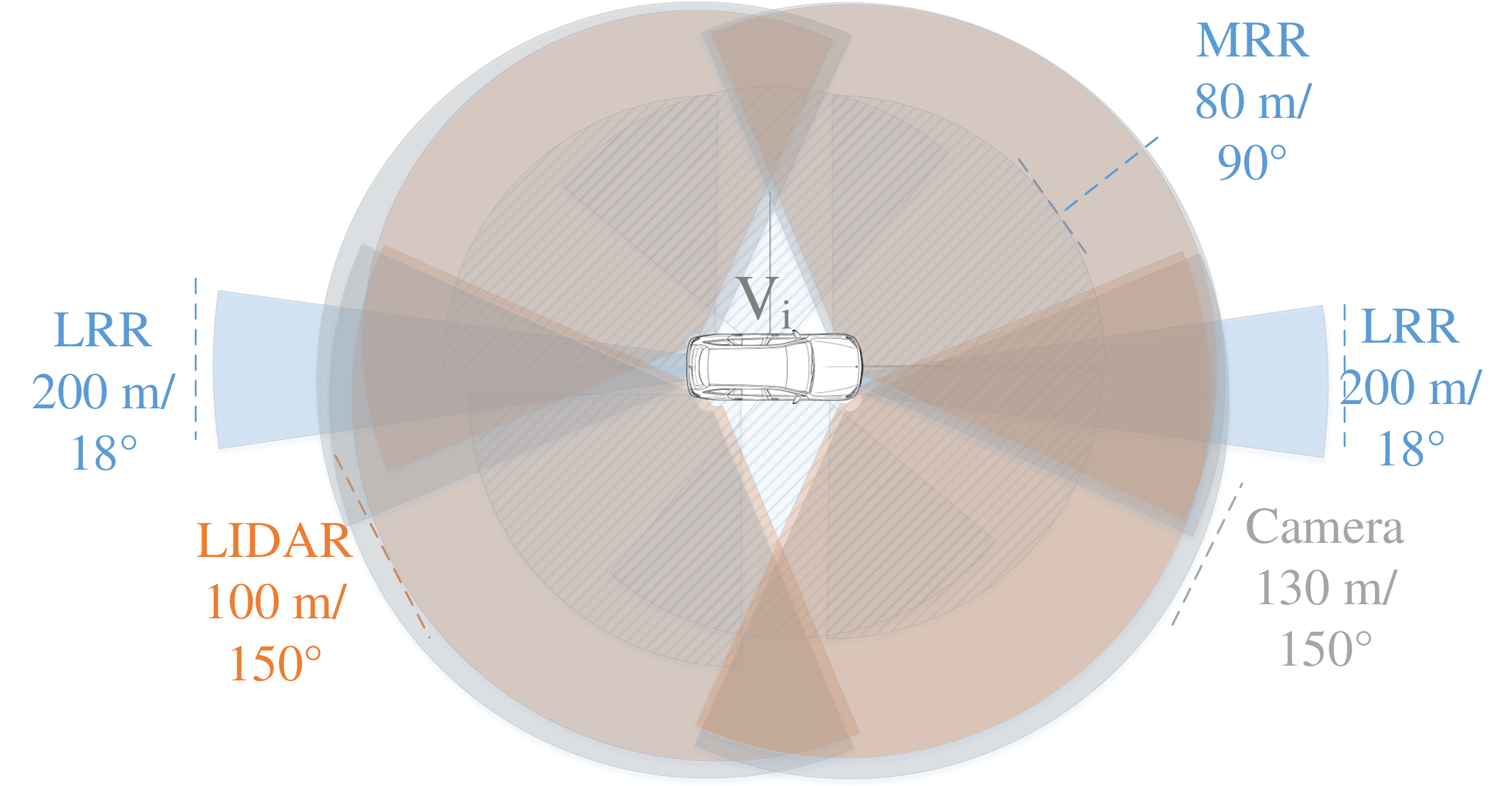}\caption{Overview of the used sensor setup and the resulting field of view.}\label{fig:sensor_view}\vspace{-0.7cm}
\end{figure} 

Predicting the future behavior of surrounding traffic participants is widely studied in literature. A good survey on motion prediction techniques in general is given in \cite{lefevre2014}. The most relevant approaches with regard to this article are the ones which Lefèvre categorizes as maneuver-based. These approaches have in common, that driving behaviors are represented through a finite set of maneuvers. In the context of highway driving the set of maneuvers is mostly reduced to the three maneuvers lane following, lane change to the left and lane change to the right. This is very similar to what is denoted as roles in the context of cooperative merging scenarios \cite{Sawade.2018}. Regarding motion prediction, this work is mainly based on the approach presented in \cite{schlechtriemen2015will, wirthmueller2019, wirthmueller2020towards}, which is introduced in more detail in \autoref{sec:sys_overview}. Other relevant works can, for example, be found in \cite{schlechtriemen2014lane, bahram2016, benterki2020artificial}.

\section{SYSTEM OVERVIEW}\label{sec:sys_overview} 
As part of this work, we implemented an extended environment model with additional v2v communication technology. \autoref{fig:ext_env_overview} shows the concept of the extended environment model implemented here. As can be seen, there are v2x-equipped (ID $V_1$) and non-equipped vehicles (ID $V_2$, ID $V_3$, ID $V_4$) on the highway. Vehicle $V_1$ sends its own position, which is estimated using the global navigation satellite system (GNSS) and (Internal Measurement Unit) IMU, to all other v2x-equipped vehicles. These other vehicles use the new information to locally extend their environment model.

Each v2x-enabled vehicle is equipped with a state-of-the-art sensor setup \cite{Dickmann.2015}.% \todo{add WAYMO Paper}. 
This sensor setup includes a front and long-rear radar (LRR) with a range of 200\,m and a beam angle of 18\degree  \,each. In addition, the vehicles are equipped with medium-range radars (MRR) as well as cameras and LIDAR's. This sensor design provides a field of view as shown in \autoref{fig:sensor_view}.

Surrounding vehicles that remain in the field of view are detected and located relative to the v2x-enabled vehicle. When the detection of the local environment of this vehicle is complete, the absolute positions are calculated. The information describing the detected objects are sent collectively as CPM in an object list to the other surrounding v2x-enabled vehicles.

%\todo{Due to implementation dependencies, in our experiment the CAM information and the CPM information were sent together as one message. ist das relevant? das ändert ja nix dran oder? würde ich weglassen} \todo{The used sensor setup considers sensor coverage by objects and measurement inaccuracies. mir ist die aussage die dahinter steht immer noch nicht ganz klar...}

The extended environment model merged the individual v2x messages into a global environment model reproduction, including the object information. As benefit, other vehicles can virtually extend their sensor range by subscribing to the extended environment model. This allows to maximize the total field of view and to detect dangerous situations earlier. This improves downstreamed motion planning with respect to comfort and safety. Fusing information from different sources is solved with an extended Kalman filter. The linking of different measured values is not the focus of our investigations. The extended Kalman filter is filled with a corresponding message ID of the respective vehicle. This structure allows an investigation of possible improvements through v2x communication in scenarios, which are difficult to handle and confusing. 

As described in \cite{Eiermann.2020}, the extended environment model serves as input for the situation analysis and maneuver planning of the ego-vehicle. Using the example of on-ramp merge, gaps in the future are identified and evaluated with a trajectory cost function algorithm. This gap detection is based on a prediction of the detected vehicles. In most cases, a Velocity-and-Lane-Keeping (VLK) model is used for prediction. In our real world data  based investigations it was found that the vehicles react very dynamically in the threading area. This led to an unsatisfactory result regarding the prediction quality. For this reason, the merge process algorithm "overlooked" many gaps that would occur in the future or recognized them too late. 

To improve the predictions, we integrated a known machine learning approach presented in detail in \cite{wirthmueller2019}. Therefore, we are able to precisely predict emerging gaps and the traffic flow. The approach has two processing stages. First, a multilyer perceptron classifier is used to determine the probability of all vehicles for all maneuvers. For vehicles which are themselves acting cooperatively, this first approximation is not necessary. Instead, the information about the upcoming maneuver can be directly derived from the recieved message. This information is afterwards used to weight different position prediction models, which are specialized on the respective maneuvers. More implementation details are outlined in \autoref{subsec:role_based_nn}.

\subsection{V2X-EXTENDED ENVIRONMENT}\label{subsec:v2x_ext_env}
The enhanced environment model is created by aggregating the environment information contained in the CAM and CPM messages from all v2x-enabled vehicles. For this purpose, it is necessary to be able to guarantee safety distance and reliability of the vehicle positions. A position determined exclusively by GNSS measurements with inaccuracies of several meters is not sufficient for this purpose \cite{Reid.2019, Tradacete.2019}. 

Based on the geometric specifications for road construction, such as road width and curvature as well as the vehicle dimensions for passenger cars, conclusions could be drawn on the requirements for the accuracy of the localization, which are described in \cite{Reid.2019}. Thus, for driving on motorways, a required lateral margin of error of 0.57\,m (0.20\,m, 95\,\%) and a longitudinal limit of 1.40\,m (0.48\,m, 95\,\%) as well as an orientation in each direction of 1.50\degree (0.51\degree, 95\,\%) were calculated.
The vehicle state $\chi_{i}^{W}$ of object $i$ in the world coordinate system $W$ is thus described with the %\todo[inline]{addatand} 
following vector \autoref{eq:state}:

\begin{equation}
  \chi_{i}^{W} = {[x_{i}^{W}, y_{i}^{W}, \psi_{i}^{W}, v_{x, i}^{W}, v_{y, i}^{W}]}^T
   \label{eq:state}
\end{equation}

Where $x_i$, $y_i$, $v_{x, i}$ , $v_{y, i}$ describe the longitidinal and lateral positions and velocities in the respective direction $\psi_i^W$ of the object. The variance of the position   $\Sigma_{\chi_{i}^{W}}$ can be calculated with the orientation angle $\psi _{i}^W$ and the rotation matrix as follows \autoref{eq:var_pos}:

\begin{equation}
\Sigma_{\chi_{i}^{W}} = \begin{bmatrix} 
\sigma _{x_i^W} \\ \sigma _{y_i^W} \\
\sigma _{\psi_{i}^W} \\
\sigma _{v_{x, i}^W} \\
\sigma _{v_{y, i}^W}\\
\end{bmatrix}  = 
\begin{bmatrix} 
\cos(\psi_{i}^W) \cdot \sigma_{x_i^V} -\sin(\psi_{i}^W) \cdot \sigma_{y_i^V} \\ 
\sin(\psi_{i}^W) \cdot \sigma_{x_i^V}+\cos(\psi_{i}^W) \cdot \sigma_{y_i^V} \\ \sigma_{\psi_i^V} \\ 
\cos(\psi_{i}^W) _{V_{i}} \cdot \sigma_{v_{x, i}^V} -\sin(\psi_{i}^W) \cdot \sigma_{v_{y, i}^V} \\ 
\sin(\psi_{i}^W) _{V_{i}}\cdot \sigma_{v_{x, i}^V}+\cos(\psi_{i}^W)\cdot \sigma_{v_{x, i}^V} \\ \end{bmatrix}
 \label{eq:var_pos}
\end{equation}

Where $V$ stands for the respective vehicle coordinate system. The CPM contains information about the surrounding vehicles that have been detected by the ego-vehicle. The measurement vector from object $i$ to object $j$ can be constructed as follows: 

\begin{equation}
\begin{aligned}
  \chi_{ij}^W = \chi_{ij}^W + C_{i}^{WV} \cdot M_{ij}^V
   \label{eq:t_beg}
\end{aligned}
\end{equation}

The measurement inaccuracies are described in the resulting covariance matrix $\Sigma_{\chi_{ij}^W}$, which describes confidence intervals from vehicle $i$ to vehicle $j$ in the world coordinate frame. The covariance matrix $\Sigma_{\chi_{ij}^W}$ indicates the measurement uncertainty in the world coordinate system caused by the sensors of vehicle $i$. To calculate this, the inaccuracies $\Sigma_{\chi_{ij}^V}$, which are recorded in the vehicle's own coordinate system $V$, must be transformed into the world coordinate system. Due to the fact that the measurement error of vehicle $i$ influences the covariance matrix 
%\todo[inline]{by transformation into the world coordinates} 
$\Sigma_{M_{ij}^W}$, the generalized error propagation law must be considered.
Thus the covariance matrix of the measurement from object $i$ to object $j$, based on the error propagation law, is calculated as follows \autoref{eq:meas_cov}:

\begin{equation}
\begin{aligned}
\Sigma_{M_{ij}^W} = J({C_{i}^{WV}, M_{ij}^V}) \cdot \Sigma_{M_{ij}^V} \cdot
 J({C_{i}^{WV}, M_{ij}^V})^T
 \label{eq:meas_cov}
\end{aligned}
\end{equation}

\begin{figure}[t!]
\centering\includegraphics[width=0.43\textwidth]{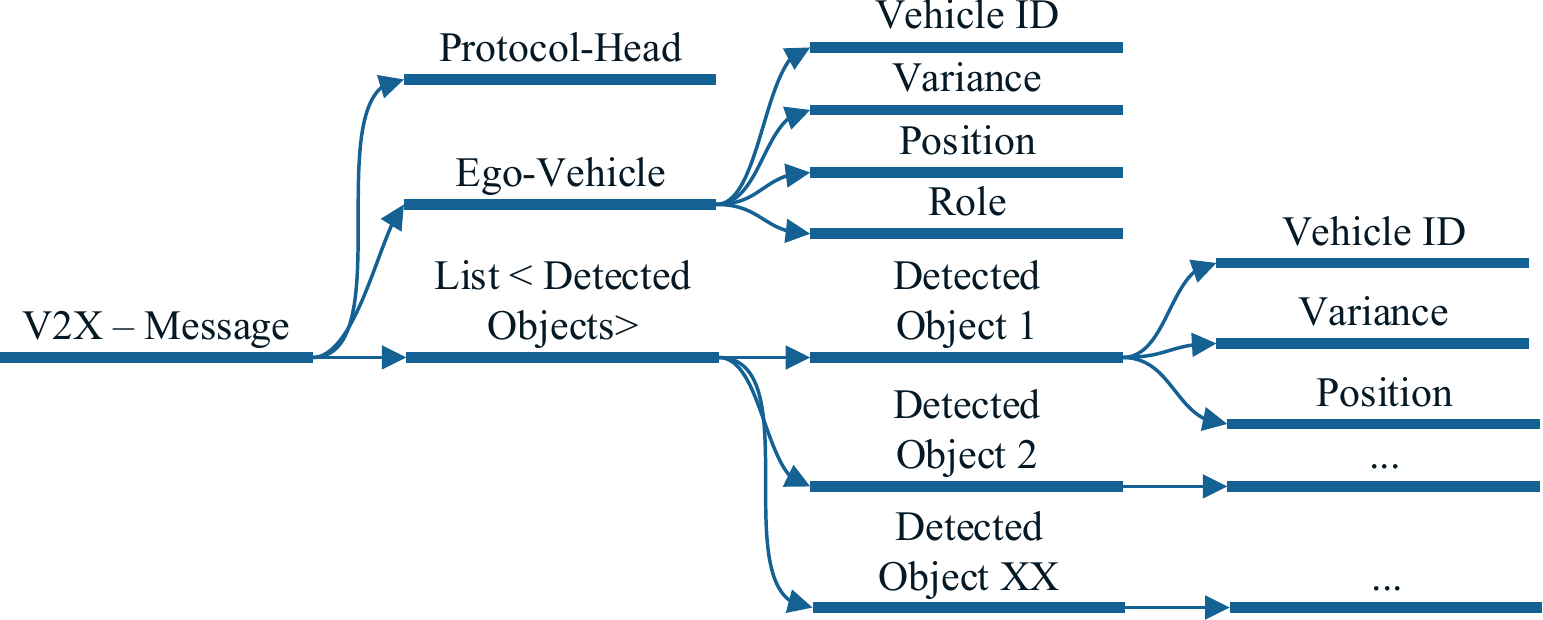}\caption{Visualization of the v2x message structure.  }\label{fig:v2x_message_struct}\vspace{-0.5cm}
\end{figure}

Where $J({C_{i}^{WV}, M_{ij}^V})$ stands for the Jacobian matrix which are the partially derived matrix of rotation matrix $C_{i}^{WV}$ multiplied by the measurement matrix $\Sigma_{M_{ij}^W}$. The vehicle position of the detected objects can thus be passed on to the extended environment model as an absolute, located state vector with associated inaccuracies.

\subsection{STRUCTURE OF THE MESSAGE}\label{subsec:struct_of_the_mess} 

The message types are implemented in the present 3D development platform PHABMACS as shown in \autoref{fig:v2x_message_struct}. As the figure shows, this work combines CAM and CPM into a single v2x message that is sent every 100 ms. The focus of our study lies not on the network load of the radio channel, but rather on evaluating the benefit of v2x communication for road users and traffic flow. For this purpose, all received v2x data must be merged to obtain an environment model that is as accurate as possible. The implementation was done with a purified Kalman filter, as described in \autoref{subsec:ext_kalman_filter}.

\subsection{EXTENDED KALMAN FILTER}\label{subsec:ext_kalman_filter}
In the extended environment model, as shown in \cite{Howard.30Sept.5Oct.2002, Karam.1315June2006, MartinSacristan.06.06.201808.06.2018, Li.24.07.201227.07.2012}, the vehicles publish their own position and the one of the detected objects. These methods create several measurement points of the ego-position distributed over the road. This concept describes a mutual relative localization of the vehicles, whereby the position accuracy can also be increased.
With the help of the Kalman filter, errors of the real measured values can be minimized and estimates for non-measurable system variables can be calculated. The filter is based on a time-discrete linear system model, which is described by mathematical equations.
In reality, the measurement and system models are usually not subject to linearity. In order to consider these in a prediction, the extended Kalman filter (EKF) was developed. Here nonlinear measurement and system models can be considered. For this purpose, the models must be linearized at the respective working point in each step of the filter using the Jacobian matrix. Thus the EKF system model can be described as follows with nonlinear equations \cite{MartinKallnik.2003} \autoref{eq:non_lin_eq}.

\begin{equation}
\begin{aligned}
\dot{x}(t)=f(x(t),u(t),w(t)) \\
y(t)=h(x(t),v(t))
 \label{eq:non_lin_eq}
\end{aligned}
\end{equation}

The nonlinear function $f$ refers to the state vector $x(t)$ and the input vector $u(t)$. The measurement vector $h$ relates the state to the measurements. The vectors $w(t)$ and $v(t)$ denote the superimposed process or measurement noise. The variance of $w(t)$ and $v(t)$ are $Q$ and $R$. 

To calculate a linearization of the models $F(t)$ and $H(t)$, the Jacobian matrix must be calculated from the nonlinear functions $f(x(t),u(t),w(t))$ and $h(x(t),v(t))$ as follows \autoref{eq:lin_jacobi_mat}:

\begin{equation}
\begin{aligned}
F(t) = \begin{bmatrix} 
	\frac{\partial f_1}{\partial x_1}  ....  \frac{\partial f_1}{\partial x_m} \\[1ex] % <-- 1ex more space between rows of matrix
 	  ................   \\[1ex]
  	\frac{\partial f_m}{\partial x_1}  ....  \frac{\partial f_m}{\partial x_m} \end{bmatrix}  ,  
H(t) = \begin{bmatrix} 
	\frac{\partial h_1}{\partial x_1}  ....  \frac{\partial h_1}{\partial x_m} \\[1ex] % <-- 1ex more space between rows of matrix
 	  ................   \\[1ex]
  	\frac{\partial h_m}{\partial x_1}  ....  \frac{\partial h_m}{\partial x_m} \end{bmatrix}
 \label{eq:lin_jacobi_mat}
\end{aligned}
\end{equation}

The construction and operation of the Kalman filter and the extended Kamlan filter are very similar. The difference is that with a Jacobian matrix the dependencies at the working point are linearized, which is a little more computationally expensive. The linearization must be recalculated for each time step. This can also be seen in \cite{MartinKallnik.2003}, where steps of the extended Kalman filter are visualized. The steps prediction and correction are the same as for the Kalman filter. However, the nonlinear functions $f$ and $h$ are linearized at the working point with the help of the Jacobian matrix. 

The vehicle information of the road users, which were dedicated and merged in the common environment model, are cyclically provided to the ego-vehicle. This includes, for example, the current position measurement, the yaw angle and its variances. The role of the vehicles are also transmitted in the v2x message. If a vehicle is not equipped with v2x technology and is detected by vehicles driving around, a path is assumed for the vehicle which follows the current track. This v2x interface to the environment model is considered an additional measurement sensor for prediction.

\subsection{ROLE-BASED BEHAVIOR PREDICTION}\label{subsec:role_based_nn}
 Within this work the prediction model introduced in \cite{wirthmueller2019} and adapted to the highD dataset \cite{krajewski2018highd} according to the remarks in \cite{wirthmueller2020towards} is used. As mentioned above, the model has two processing stages. In the first stage, a multilayer perceptron, which is a simple form of a neural network, is used to predict upcoming maneuvers. The network comprises of a single hidden layer with 27 neurons. The output of this first stage are for each vehicle in question the probabilities for the three maneuvers lane following, lane change to the right and to the left. The estimation is based on a combination of 23 different features. Thereby, the distances and relative velocities to the lane markings and surrounding vehicles as well as the type of the lane marking have a substantial impact.

\begin{figure}[t!]
\centering\includegraphics[width=0.38\textwidth]{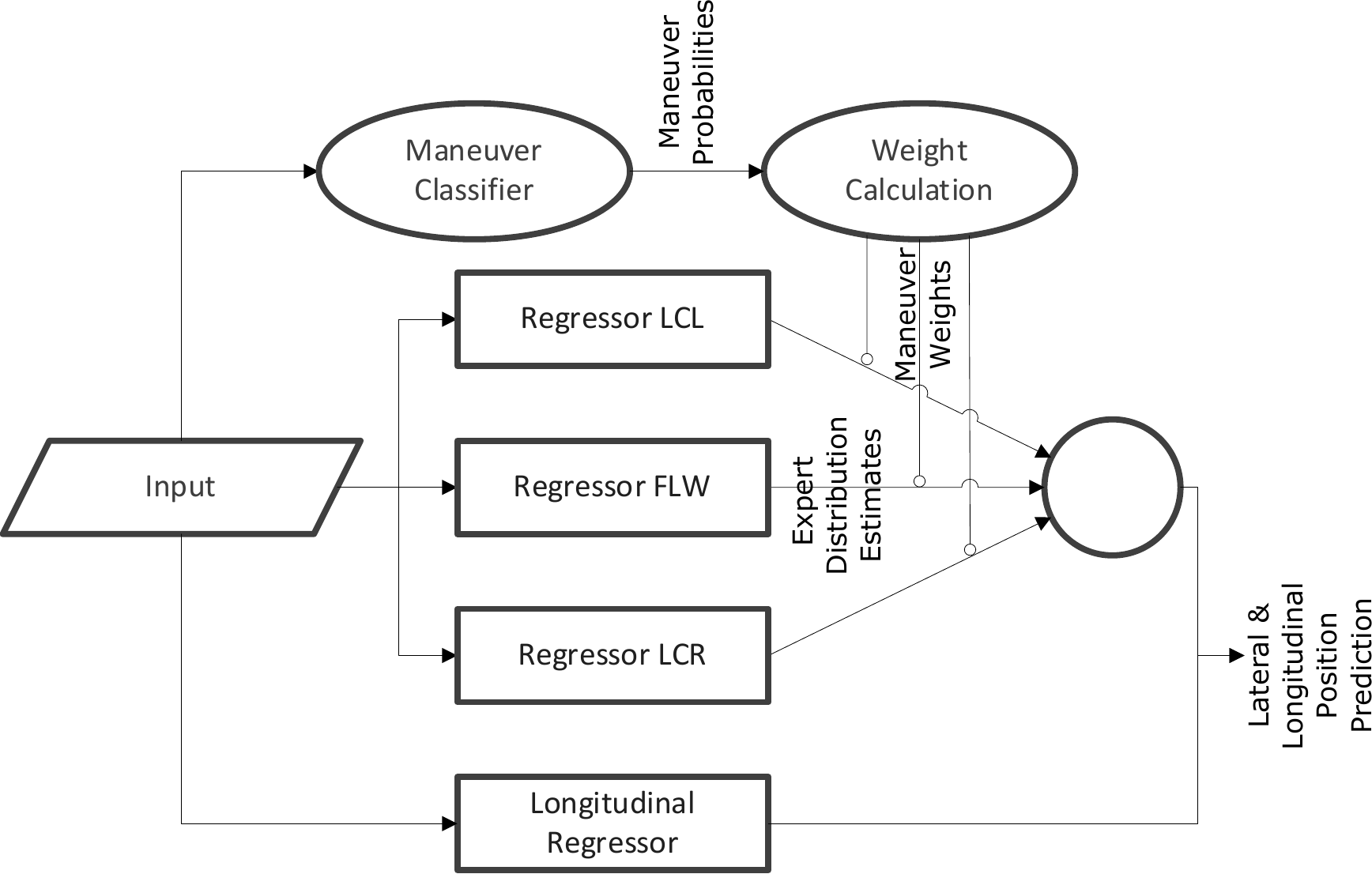}\caption{Visualization of the entire prediction workflow.}\label{fig:pred_architecture}\vspace{-0.5cm}
\end{figure}

The second stage consists of three expert position prediction models (one for each maneuver) for the lateral direction and a single one for the longitudinal one. All position predictors are modeled as Gaussian mixture models (GMM). The GMMs are used in an Gaussian mixture regression (GMR) manner. Simply put, the GMMs were trained in the combined input (measurements and prediction horizon) and output (position) space. Hence, it is possible to condition them with the current measurement values to obtain a multimodal probability distribution in the output space. The outputs of the three expert models are weighted with the maneuver probabilities and summed up. Thus, for each lateral and longitudinal direction a spatial probablity density is obtained. These distributions can then be used to estimate the most likely position of the vehicle at a specified point in the future. The overall prediction architecture is visualized in \autoref{fig:pred_architecture}.

%\todo[inline]{[folgende Punkte habe ich zur Vereinfachung weggelassen]}
%
%\begin{itemize}
%\item priors multiplizieren
%\item normalisieren
%\item unterscheidung zwischen mit und ohne vordermann 
%\item auf highD trainiert --> damit ist nicht auszuschließen, dass die jetzt zur validierung verwendeten szenarien in den trainingsdaten enthalten waren
%\item das spielt aber nur eine geringe rolle, da das prädiktionsmodell auch darauf ausgelegt ist mit realen sensorinputs umzugehen --> fahrzeuge die weiter als XX\,m weg sind wurden daher ausgenommen --> mit einem retrain könnte man die performance somit in etwa gleichem mase verbessern
%\end{itemize} 

\section{IMPLEMENTATION AND SIL TESTING}\label{sec:imp_and_sil_testing}
The 3D simulation environment PHABMACS was used to implement the extended environment model and the neural network connection. It includes extensive driving physics simulations and toolboxes for v2x network transmission. Furthermore, the simulation environment is coupled to the real world highD data set \cite{krajewski2018highd} provided by RWTH Aachen. This means that the vehicle movements in the simulation exactly replicate those of a real shot of the A4 near Cologne motorway intersection. By a manipulated coincidence, vehicles are equipped with a v2x skill. This enables a real penetration of communicating (POC) vehicles. Different penetration percentages are selected in the tests to be able to make a statement about the necessary communication penetration. The threading vehicles are abstracted from the data set and controlled by an EgoMergeSkill. This skill finds a suitable gap using the extended environment model and plans a drivable trajectory for the on-ramp merging process. The exact procedure of this skill is described in \cite{Eiermann.2020}.

To convert the simulation into a software in the loop (SIL) test environment, an interaction skill was implemented. This checks if an interaction with the merging vehicle exists. If this is the case, the vehicle that actually follows the real vehicle data is transformed into a PHABMACS simulated vehicle with adaptive cruise control (ACC). This is a requirement for a function test, which is based on real vehicle data in traffic.

%In order to get our software in the loop, an additional skill was implemented for the \todo{real-data vehicles - was meinst du damit?}. This skill checks if there is an interaction with the threading vehicle. This is useful, for example, when the threading vehicle and the merging vehicle come very close to each other, or when they are prevented from free movement by another vehicle. If this is the case, the real vehicle becomes a \todo{simulated vehicle - verstehe ich nicht...}. This is a prerequisite for a functional test, which is based on real driving profiles recorded in traffic. 

The skills and the simulation environment are installed on an i9 Dell Precision Workstations. The execution time is in real time so that the required computing power is very limited. 
For a calculation of a trajectory, i.\,e\,. network and regressor, the processor needs a computing time of 100\,ms. Furthermore, the computing time of the environment model is measured by the Kalman filter at approx. 300\,ms. on average. This is of course always dependent on the number of measuring points to be merged and linear to the number of vehicles considered in the environment model. 

\begin{figure}[t!]
\centering\includegraphics[width=0.33\textwidth]{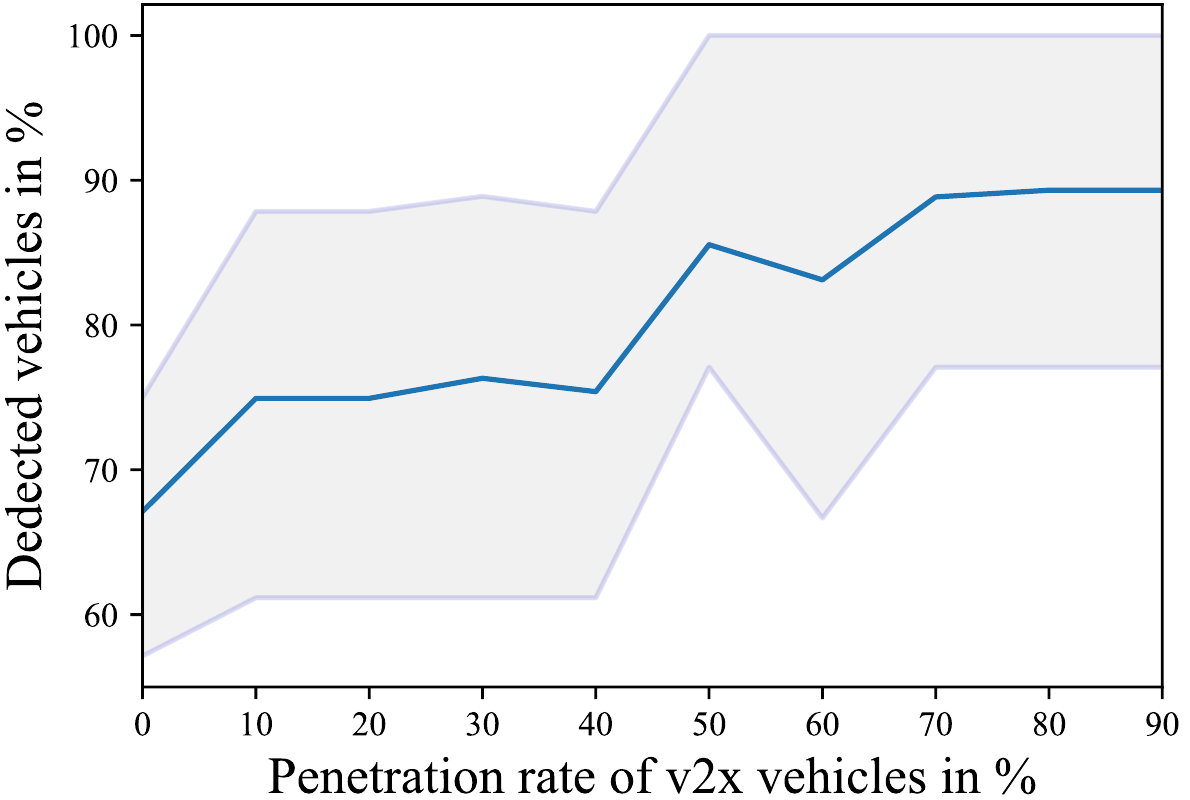}\caption{Detected vehicles during on-ramp merging in dependence of the v2x penetration rate. The quartiles (0.25, 0.75) are marked in grey}\label{fig:detect_obj}\vspace{-0.7cm}
\end{figure} 

In our investigations we focus on the additional gain of the extended environment model and the subsequent neural network prediction. Furthermore, the v2x penetration is to be evaluated in relation to the threading process.  
As our main Key Performance Indicator, we have taken into account the number of all detected objects as well as acceleration and spine, which can be detected by misinterpreting the prediction in the threading maneuver. Furthermore, the prediction error to an existing constant velocity model is of course also examined.

\section{EVALUATION AND RESULTS}\label{sec:eval_and_results}
The benefit of the extended environment model can be seen from the number of covered objects. In the ideal case, the extended environment model includes all vehicles in the environment of the ego-vehicle. Thus, the trajectory planning can consider while calculating an ideal trajectory for the ego-vehicle. Furthermore, the position error of the objects has a considerable impact on the quality of the environment model. The more precisely a vehicle can be located, the better the free space for possible maneuvers can be considered during the trajectory planning. These two main quality criteria were examined using the example of a threading process. 

\subsection{DETECTED OBJECTS}\label{subsec:det_obj}
Especially during threading operations, it is important to get as much information as possible about the traffic, the vehicles and their speeds of the main roadway. As mentioned in the system overview \autoref{sec:sys_overview}, the simulation environment PHABMACS was used to play back real world vehicle data in order to perform the described evaluations. At the beginning of the used scenario, we can set a v2x penetration rate and therefore parameterize the scenario. Based on a manipulated random calculation that maps the v2x penetration rate, real vehicles either equipped with a v2x skill or not. Vehicles that have this skill send CAMs and CPMs to the extended environment model. For this purpose they are equipped with the sensor setup described in \autoref{sec:sys_overview}. Of course they follow their trajectory from the real world data set. All other randomly selected vehicles, which do not have v2x capability, follow their recorded trajectory from the real world data set as well.
In our studies we have investigated the number of detected objects in relation to the equipment rate of the surrounding traffic. For this purpose, we ran several simulations using different penetration rates from 0\,\% to 90\,\%. For each simulation time step, the number of detected objects in the extended environment model was logged. \autoref{fig:detect_obj} shows the results of the investigations.

The number of detected objects was taken in relation to all vehicles located within a 400\,m radius around the ego-vehicle. During the test, the number of vehicles in the surrounding environment of the ego-vehicle ranged from 3 to 9.

%% From KAY: 
This range of surrounding vehicles results from one specific scenario in the highD dataset. In order to validate these numbers against general conditions in real world, we conducted a study using real world fleet data \cite{wirthmueller2019}. A statistical analysis of data acquired from radar and camera sensors of Daimler vehicles showed that this range is a sensible assumption for on-ramp merging situations, given this type of scenario and day time.

%\todo{---------------KAY:  Wie KNFE Daten zeigen, ist dies bei Einfädelvorgängen eine heufig detektierte Anzahl an Objekten, womit das verwendete Sensor-Setup (RADAR) stochastisch validiert wurde  - ILJA}

%Thus, a relative evaluation of the environmental representation can be conducted.
As can be seen, with a penetration rate of 90\,\% to 70\,\% on average almost 90\,\% of all objects are included in the extended environment model. From a penetration rate of 60-50\,\% on, there is a small reduction of the detected objects to about 80\,\%.
At a penetration rate of 40-10\,\%, a gradation is to be noted. In comparison, only 70\,\% vehicles in the surrounding model are shown here. These results were also confirmed in \cite{schiegg2019object, schiegg2019itsc}. 
% maybe use one ! 
%\todo{(TODO Florian Schiegg \cite{schiegg2019object, schiegg2019itsc} – PaPer verweis)}
If there is no v2x-capable vehicle in the threading lane, the ego-vehicle detects 68\,\% of all road users. This is a comparatively high ratio, as the motorway access lane does not take into account any structural shading by e.g. crash barriers or static objects such as bushes on the side of the road. Furthermore, the entry lane runs parallel to the main road for a very long time, so that objects are already detected there. If the highway feeder road would lead to the threading lane in a curve, the number of detected objects would be much lower.

\subsection{EXTENDED ENVIRONMENT}\label{subsec:ext_env_errror}

\begin{figure}[t!]
\centering\includegraphics[width=0.33\textwidth]{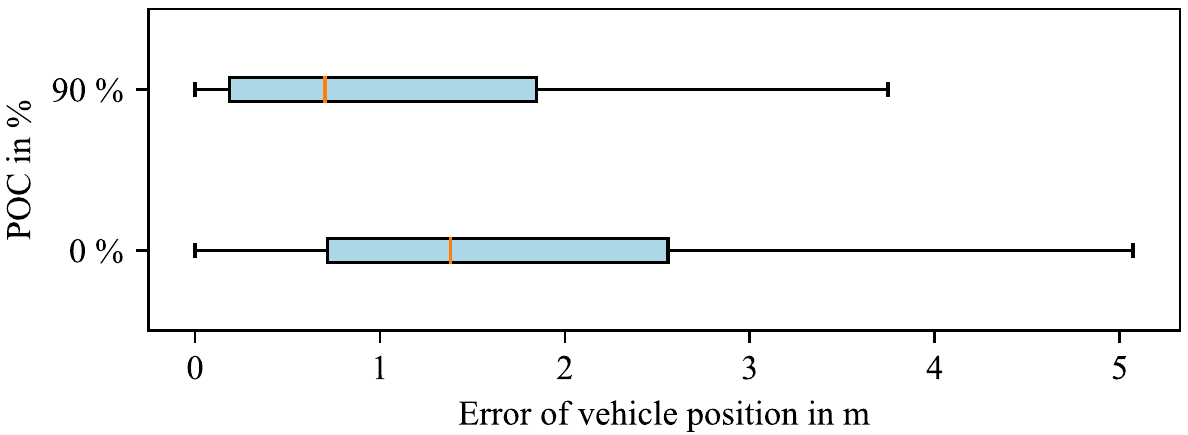}\caption{Comparison of the localization error of vehicles at 0\,\% v2x penetration rate (POC) (ego sensor only) and 90\,\% penetration rate}\label{fig:error_ext_env}\vspace{-0.6cm}
\end{figure}

In this experiment, were compared the position accuracy resulting from direct sensor measurement with the position accuracy resulting from the additional employment of v2x. As depicted in \autoref{fig:error_ext_env}, the accuracy could be improved to 0.7\,m at 90\,\% v2x penetration, compared to 1.4\,m without v2x. Our rational for creating a fair setup for comparison in simulation is in order. In real world, the determination of the position of surrounding vehicles using sensors is subject to a delay caused by object detection technics. In case of radar sensors, this delay may be about 150\,ms or higher. For the communication delay, we can assume a lower value of about 50\,ms in real world. Both delays influence the position error in a different way. The communication delay impacts the absolute positioning and is, thus, sensitive to the absolute velocity of a vehicles. Given a velocity of about 100\,km/h at highways, this delay causes a longitudinal position error of about 1.38\,m. The sensor delay is sensitive to the relative velocity between the ego-vehicle and the detected object. The prediction model presented in \autoref{subsec:role_based_nn} reduces the v2x related positioning error only, which would bias the comparison. We, therefore, decided to omit a sensor delay in simulation and to set the communication delay to 50\,ms. 
%Optional kann man noch schreiben:
%The object sensor in simulation is configured with noise on the position using a Gaussian error of \todo{1.5  … hier die werte eintragen und die referenz wo der wert herkommen… }. The noise of the absolute positioning of the vehicles communicated via v2x is negligible, assumed that capable GNSS correction data is available.  
\subsection{ROLE-BASED BEHAVIOR PREDICTION}\label{subsec:rol_based_nn_pred_error}

\begin{figure}[t!] 
\centering\includegraphics[width=0.35\textwidth]{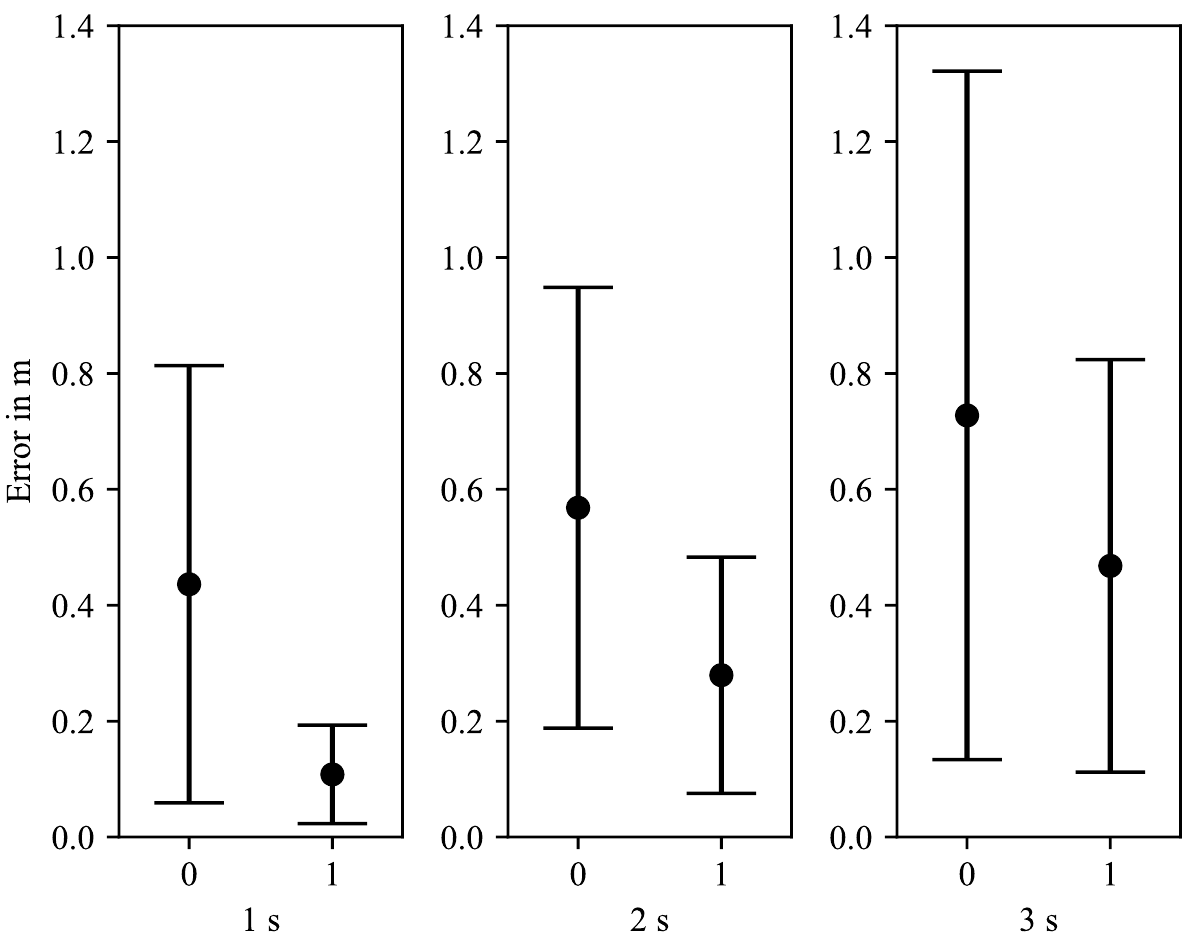}\caption{Lateral prediction error of velocity-and-lane-keeping (VLK) (0) and role-based behavior model (RBB) (1) over the prediction horizon of 1 to 3\,s. The plot shows the average values (depicted as dot) and the maximum and minimum values.}\label{fig:error_pred}\vspace{-0.5cm}
\end{figure}

For a successful merging process, the prediction quality of surrounding objects is also an important factor and can significantly improve the trajectory planning. In this work, the role-based motion prediction approach is investigated in comparison to a constant velocity model. \autoref{fig:error_pred} shows the lateral prediction errors of the different approaches for prediction times ranging from 1\,s to 3\,s. As can be seen, the role based position prediction approach produces estimates whith significantly smaller errors than the VLK model. Whereas in the original study \cite{wirthmueller2020towards} the median prediction error is given, we decided for the average errors. This is motivated, as we recognized that the median values are more subjected to noise because of our relatively small sample size. Hence we are refering to \cite{wirthmueller2020towards} for an in depth evaluation of the prediction module. Nevertheless, our evaluations show, that the RBB shows up a significantly improvement with respect to the lateral prediction error over all time horizons. 

The longitudinal error is very similar between the two approaches. At a prediction horizon of 1\,s it amounts approximately 2\,m. At 2\,s about 2.5\,m and at 3\,s about 3\,m. But also here, the RBB approach has to be favored against the VLK model at all times, showing improved average errors approximately ranging from 0.25 to 0.5\,m. 

Besides, it is remarkable that the RBB has learned the right-hand drive law holding true in Germany. This can be concluded from single situations where vehicles are predicted to perform a lane change to the right if their right hand lane is not occupied.

As stated above an improved prediction module contributes to improve the planning of threading trajectories e.\,g. in a cooperative maneuver where the partner changes lanes and can therefore plan his trajectory better.  

\subsection{TRAJECTORY PLANING WITH EXTENDED ENVIRONMENT}\label{subsec:traj_ext_env}

In the following, we investigated the overall system with the modules extended environment and role-based neural network prediction. For this purpose, several driving parameters and safety parameters were recorded, during on-ramp merging processes. Among the driving parameters we logged the longitudinal and lateral acceleration and the speed. Among the safety parameters we logged the distance to vehicles on the same lane $dist$ and the Time To Collision $ttc$. These parameters are listed in \autoref{tab:eval}. The driving parameters are the average value of the recorded measured values. The safety-relevant parameters are shown here with the minimum value that occurred.
The simulation threading processes were performed once without v2x (extended environment model) and without a role-based prediction.  In a second step, we added the role-based prediction and the v2x capability and thus an extended environment model. We did this once with 50\,\% and 80\,\% v2x penetration. 

  \begin{table}[!ht]
  \caption{on-ramp merging parameter}%\todo{Title - je spalte immer gleiche zahl an nachkommastellen!}}
	\label{tab:eval}
 	\centering
 	\begin{tabular}{|c|c|c|c|c|c|}
 		\hline
		v2x rate & \multicolumn{3}{c|}{driving parameters} & \multicolumn{2}{c|}{safety parameters} \\
		& $\overline{a_{x}}$ & $\overline{a_{y}}$ & $\overline{v}$ & $\min(dist)$ & $min(ttc)$ \\
		$[\%]$ & $[\nicefrac{m}{s^2}]$ & $[\nicefrac{m}{s^2}]$ & $[\nicefrac{m}{s}]$ & $[m]$ & $[s]$ \\
 		\hline
		0 & 1.25 & 0.23 & 19.60 & 12.08 & 3.01 \\
 		\hline
		50 & 1.13 & 0.17 & 19.64 & 13.94 & 3.30 \\
 		\hline
		80 & 0.92 & 0.13 & 19.06 & 23.92 & 5.29 \\
 		\hline
 		\end{tabular}\vspace{-0.4cm}
 \end{table}
 
As can be seen, the accelerations are smaller with increasing v2x penetration. This means that, as assumed at the beginning, an extended environment model helps to plan better trajectories. Thus, trajectories with less acceleration can be found. The longitudinal accelerations have almost halved from the output value without v2x. This is an indication that the trajectory planner performs the lane change more smoothly ensuring increased comfort.The longitudinal acceleration value is minimized by about 0.33$\,\nicefrac{m}{s^2}$. This is an hint that the longitudinal path planning with the extended environment model can take more surrounding vehicles into account.
%This is an \todo{indication} that longitudinal path planning with the extended environment model can take more surrounding vehicles into account. 
Thus, the trajectory planner can control the gap at an early stage and adjust the speed. The threading speed has changed very little over the different measurements. At the minimum distances the advantages of an extended environment model is clearly visible. Here the distance to the surrounding vehicles almost doubled during the entire threading process. 
%\todo{This is remarkable.} 
The Time To Collision behaves in the same way. 
%These are \todo{very good results} and an essential finding of the investigation. 
Since the acceleration values change only slightly (0.33\,$\nicefrac{m}{s^2}$) from 0\,\% v2x to 80\,\% v2x, this shows that the trajectory planner starts the acceleration early enough and can plan it very far in advance due to the extended environment model. Thus, the vehicle is brought to the right speed and position for threading very early.  

\section{SUMMARY AND OUTLOOK}\label{sec:conclusion}

In this paper we present the benefits of v2v communication especially for the threading process. Using CAM and CPM extended environment models can be created, which can support the ego trajectory planning. We found that a v2x vehicle penetration of 10\,\% can be used to cover 75\,\% of the environment. With increasing number of v2x participants, of course, the penetration is increasing. 
With role-based prediction algorithms, which are based on neural networks, a very accurate prediction of maneuvers and future position can be made. A median prediction error of less than 0.4\,m in lateral and 3\,m in longitudinal direction for a prediction horizon of 3\,s can be achieved. Thus, gaps in flowing traffic can be detected very well. 

The overall integration also shows that the coupling of an extended environment model and a role-based prediction has a very positive effect on the comfort and safety during threading. 
In the future, these advantages could be supplemented with suitable Road-Side-Units (RSU) for highly frequented motorway entrances. This allows an almost ideal extension of the extended environment model, so that vehicles would benefit in the best possible way in terms of safety and comfort. 
Further work will now investigate how v2x cooperation maneuvers can be used to improve the on-ramp merging.

\section*{ACKNOWLEDGMENTS}\label{sec:acknol}
The design, implementation, and testing activities leading to this contribution were partially realized as a part of research project IMAGinE “Intelligent Maneuver Automation – cooperative hazard avoidance in real-time”, thankfully funded by the German Federal Ministry for Economic Affairs and Energy.

\addtolength{\textheight}{-6cm}   % This command serves to balance the column lengths

\bibliographystyle{ieeetr}
%\nocite{*}

\bibliography{bib_iccp2020}

\interlinepenalty=10000
\begin{thebibliography}{10}

\bibitem{Massow.2018}
K.~Massow, I.~Radusch, and Y.~Du, ``A rapid prototyping environment for
  cooperative advanced driver assistance systems,'' {\em Journal of Advanced
  Transportation}, vol.~2018, p.~2586520, 2018.

\bibitem{EuropeanTelecommunicationsStandardsInstitute.2013}
{European Telecommunications Standards Institute}, ``{EN} 302 637-2 - v1.3.1 -
  intelligent transport systems ({ITS}); vehicular communications; basic set of
  applications; part 2: Specification of cooperative awareness basic service,''
  2013.

\bibitem{SAEInternational.2016}
{SAE International}, ``Dedicated short range communications ({DSRC}) message
  set dictionary{\texttrademark},'' 2016.

\bibitem{Dixit.2018}
S.~Dixit, S.~Fallah, U.~Montanaro, M.~Dianati, A.~Stevens, F.~Mccullough, and
  A.~Mouzakitis, ``Trajectory planning and tracking for autonomous overtaking:
  State-of-the-art and future prospects,'' {\em Annual Reviews in Control},
  vol.~45, pp.~76--86, 2018.

\bibitem{Sawade.2018}
O.~Sawade, M.~Schulze, and I.~Radusch, ``Robust communication for cooperative
  driving maneuvers,'' {\em IEEE Intelligent Transportation Systems Magazine},
  vol.~10, no.~3, pp.~159--169, 2018.

\bibitem{Eiermann.2020}
L.~Eiermann, S.~B. Bunk, and I.~Radusch, ``Cooperative automated lane merge
  with role-based negotiation,'' 2020.
\newblock Will be published at {IEEE IV} 2020. Please reach out, if you are
  interested in the full text!

\bibitem{Becker.2000}
J.~C. Becker, ``Adaptive information filter for the fusion of data from the
  object-detecting sensors of an autonomous vehicle,'' {\em IFAC Proceedings
  Volumes}, vol.~33, no.~9, pp.~247--252, 2000.

\bibitem{Becker.58Oct.1999}
J.~C. Becker, ``Fusion of data from the object-detecting sensors of an
  autonomous vehicle,'' in {\em Proceedings 199 IEEE/IEEJ/JSAI International
  Conference on Intelligent Transportation Systems (Cat. No.99TH8383)},
  pp.~362--367, IEEE, 5-8 Oct. 1999.

\bibitem{Dickmann.2015}
J.~Dickmann, N.~Appenrodt, J.~Klappstein, H.-L. Bloecher, M.~Muntzinger,
  A.~Sailer, M.~Hahn, and C.~Brenk, ``Making bertha see even more: Radar
  contribution,'' {\em IEEE Access}, vol.~3, pp.~1233--1247, 2015.

\bibitem{lefevre2014}
S.~Lef{\`e}vre, D.~Vasquez, and C.~Laugier, ``A survey on motion prediction and
  risk assessment for intelligent vehicles,'' {\em ROBOMECH journal}, vol.~1,
  no.~1, p.~1.
\newblock Nature Publishing Group, 2014.

\bibitem{schlechtriemen2015will}
J.~Schlechtriemen, F.~Wirthmueller, A.~Wedel, G.~Breuel, and K.-D. Kuhnert,
  ``When will it change the lane? {A} probabilistic regression approach for
  rarely occurring events,'' in {\em 14th Intelligent Vehicles Symposium (IV)},
  pp.~1373--1379, IEEE, 2015.

\bibitem{wirthmueller2019}
F.~Wirthm{\"u}ller, J.~Schlechtriemen, J.~Hipp, and M.~Reichert, ``Teaching
  vehicles to anticipate: A systematic study on probabilistic behavior
  prediction using large data sets,'' {\em Transactions on Intelligent
  Transportation Systems (T-ITS)}, pp.~1--16.
\newblock IEEE, 2020.

\bibitem{wirthmueller2020towards}
F.~Wirthm{\"u}ller, J.~Schlechtriemen, J.~Hipp, and M.~Reichert, ``Towards
  incorporating contextual knowledge into the prediction of driving behavior,''
  {\em arXiv preprint arXiv:2006.08470}, 2020.

\bibitem{schlechtriemen2014lane}
J.~Schlechtriemen, A.~Wedel, J.~Hillenbrand, G.~Breuel, and K.-D. Kuhnert, ``A
  lane change detection approach using feature ranking with maximized
  predictive power,'' in {\em 2014 IEEE Intelligent Vehicles Symposium
  Proceedings (IV)}, pp.~108--114, IEEE, 2014.

\bibitem{bahram2016}
M.~Bahram, C.~Hubmann, A.~Lawitzky, M.~Aeberhard, and D.~Wollherr, ``A combined
  model-and learning-based framework for interaction-aware maneuver
  prediction,'' {\em Transactions on Intelligent Transportation Systems},
  vol.~17, no.~6, pp.~1538--1550.
\newblock IEEE, 2016.

\bibitem{benterki2020artificial}
A.~Benterki, M.~Boukhnifer, V.~Judalet, and C.~Maaoui, ``Artificial
  intelligence for vehicle behavior anticipation: Hybrid approach based on
  maneuver classification and trajectory prediction,'' {\em IEEE Access},
  vol.~8, pp.~56992--57002.
\newblock IEEE, 2020.

\bibitem{Reid.2019}
T.~G.~R. Reid, S.~E. Houts, R.~Cammarata, G.~Mills, S.~Agarwal, A.~Vora, and
  G.~Pandey, ``Localization requirements for autonomous vehicles,'' {\em SAE
  International Journal of Connected and Automated Vehicles}, vol.~2, no.~3,
  2019.

\bibitem{Tradacete.2019}
M.~Tradacete, {\'A}.~S{\'a}ez, J.~F. Arango, C.~{G{\'o}mez Hu{\'e}lamo},
  P.~Revenga, R.~Barea, E.~L{\'o}pez-Guill{\'e}n, and L.~M. Bergasa,
  ``Positioning system for an electric autonomous vehicle based on the fusion
  of multi-gnss rtk and odometry by using an extented kalman filter,'' in {\em
  Advances in Physical Agents} (R.~{Fuentetaja Piz{\'a}n}, {\'A}.~{Garc{\'i}a
  Olaya}, M.~P. {Sesmero Lorente}, J.~A. {Iglesias Mart{\'i}nez}, and
  A.~{Ledezma Espino}, eds.), vol.~855 of {\em Advances in Intelligent Systems
  and Computing}, pp.~16--30, Cham: {Springer International Publishing}, 2019.

\bibitem{Howard.30Sept.5Oct.2002}
A.~Howard, M.~J. Matark, and G.~S. Sukhatme, ``Localization for mobile robot
  teams using maximum likelihood estimation,'' in {\em IEEE/RSJ International
  Conference on Intelligent Robots and System}, pp.~434--439, IEEE, 30 Sept.-5
  Oct. 2002.

\bibitem{Karam.1315June2006}
N.~Karam, F.~Chausse, R.~Aufrere, and R.~Chapuis, ``Cooperative multi-vehicle
  localization,'' in {\em 2006 IEEE Intelligent Vehicles Symposium},
  pp.~564--570, IEEE, 13-15 June 2006.

\bibitem{MartinSacristan.06.06.201808.06.2018}
D.~Martin-Sacristan, S.~Roger, P.~Spapis, A.~Kousaridas, and C.~Zhou,
  ``Low-latency v2x communication through localized mbms with local v2x servers
  coordination,'' in {\em IEEE International Symposium on Broadband Multimedia
  Systems and Broadcasting}, pp.~1--8, 2018.

\bibitem{Li.24.07.201227.07.2012}
H.~Li and F.~Nashashibi, ``Multi-vehicle cooperative localization using
  indirect vehicle-to-vehicle relative pose estimation,'' in {\em 2012 IEEE
  International Conference on Vehicular Electronics and Safety (ICVES 2012)},
  pp.~267--272, IEEE, 2012.

\bibitem{MartinKallnik.2003}
{Martin Kallnik}, ``Sensordatensynchronisation zum {A}ufbau eines
  {N}avigationskalmanfilters f{\"u}r die mobile {R}obotik,'' 2003.

\bibitem{krajewski2018highd}
R.~Krajewski, J.~Bock, L.~Kloeker, and L.~Eckstein, ``The high{D} dataset: A
  drone dataset of naturalistic vehicle trajectories on german highways for
  validation of highly automated driving systems,'' in {\em 21st International
  Conference on Intelligent Transportation Systems (ITSC)}, pp.~2118--2125,
  IEEE, 2018.

\bibitem{schiegg2019object}
F.~A. Schiegg, I.~Llatser, and T.~Michalke, ``Object detection probability for
  highly automated vehicles: An analytical sensor model.,'' 2019.

\bibitem{schiegg2019itsc}
F.~A. {Schiegg}, N.~{Brahmi}, and I.~{Llatser}, ``Analytical performance
  evaluation of the collective perception service in c-v2x mode 4 networks,''
  in {\em 2019 IEEE Intelligent Transportation Systems Conference (ITSC)},
  pp.~181--188, 2019.

\end{thebibliography}

\end{document}